%Paper: hep-ph/9410230
%From: konig@osiris.phy.uqam.ca (Heinz Konig)
%Date: Wed, 5 Oct 94 21:53:29 -0400

\magnification=\magstep1\pageno=1
\overfullrule=0pt
\hsize=15.4truecm
\line{\hfil UQAM-PHE-94/08}
\vskip2cm
\centerline{\bf FLAVOUR CHANGING TOP QUARK DECAY WITHIN}
\centerline{\bf THE MINIMAL SUPERSYMMETRIC STANDARD MODEL}
\vskip2cm
\centerline{G. COUTURE, C. HAMZAOUI AND H. K\"ONIG
\footnote*{email:couture, hamzaoui, konig@osiris.phy.uqam.ca}}
\centerline{D\'epartement de Physique}
\centerline{Universit\'e du Qu\'ebec \`a Montr\'eal}
\centerline{C.P. 8888, Succ. Centre Ville, Montr\'eal}
\centerline{Qu\'ebec, Canada H3C 3P8}
\vskip2cm
\centerline{\bf ABSTRACT}\vskip.2cm\indent
We present the results of the
gluino and scalar quarks contribution to
the flavour changing top quark decay into a
 charm quark and a photon, gluon or a $Z^0$\ boson
within the minimal supersymmetric standard
model. We include the mixing of the scalar partners
of the left and right handed top quark.
This mixing has several effects, the most
important of which are to greatly enhance the c Z
decay mode for large values of the soft SUSY breaking
scalar mass $m_S$\ and to give rise to a GIM--like suppresion
in the c $\gamma$\ mode for certain combinations of
parameters.
\vskip.5cm
\centerline{ September 1994}
\vfill\break
Recent experimental evidence of the top quark [1]
makes its rare decay modes a promising test ground for the
standard model (SM) and physics beyond the SM.
The flavour--changing decay mode of the top quark was
calculated within the SM in [2-5]
and shown to be far away from experimental reach; this
makes it an excellant probe for models beyond
the SM. Two-Higgs-doublet models (THDM)were
considered in [6,7], where it was shown that the decay
rate is enhanced by several (3--4) orders of magnitude.
Recently [8], the $t\rightarrow
cV$ decay was considered
within the minimal supersymmetric SM (MSSM) and the authors obtained the
same enhancement as in the THDM's. However they did
not include the mixing of the scalar partners of the
left and right handed top quark, they omitted one diagram
in the $c~g$ decay mode and their current was not gauge invariant.
\hfill\break\indent
In this paper we present the QCD loop corrections to the
 $t\rightarrow c V$\ decay in the MSSM with gluinos and scalar quarks
running on the loop, as shown in fig.1. Throughout
the calculation we neglect all quark masses besides
the top quark mass and include
the mixing of the scalar partners of
the left and right handed
top quark, which is proportional to the top quark
mass.
\hfill\break\indent
In supersymmetric QCD it was shown that there occur flavour
changing strong interactions between the gluino, the left
handed quarks and their supersymmetric scalar partners,
whereas the couplings of the gluino to the right handed
quarks and their partners remains flavour diagonal [10-16].
Since the mixing of $\tilde t_L$\ and $\tilde t_R$\
is proportional to the top quark mass we have to include the
full scalar top quark matrix which is given by [9]:
$$M^2_{\tilde t}=\left(\matrix{m^2_{\tilde t_L}
+m_{\rm top}^2+0.35D_Z^2&
-m_{\rm top}(A_{\rm top}+\mu\cot\beta)\cr-m_{\rm top}
(A_{\rm top}+\mu\cot\beta)&
m^2_{\tilde t_R}+m^2_{\rm top}+0.16D_Z^2\cr}\right)\eqno(1)$$
where $D_Z^2=m_Z^2\cos 2\beta$,
$m^2_{\tilde t_{L,R}}$\ are soft SUSY breaking masses,
$A_{\rm top}$\ is a trilinear
scalar interaction parameter and $\mu$\ is the supersymmetric
mass mixing term of the Higgs bosons. The mass eigenstates
$\tilde t_1$\ and $\tilde t_2$\ are related to the current
eigenstates $\tilde t_L$\ and $\tilde t_R$ by:
$$\tilde t_1=cos\Theta_t\tilde t_L+\sin\Theta_t\tilde t_R\qquad
\tilde t_2=-\sin\Theta_t\tilde t_L+\cos\Theta_t\tilde t_R\eqno(2)$$
 In the following we take
$m_{\tilde t_L}=m_{\tilde t_R}=m_S=A_{\rm top}$\
(global SUSY).
The gluino mass $m_{\tilde g}$\ is a free parameter,
which in general is supposed to be larger than 100 GeV, although
there is still the possibility of a small gluino mass window in
the order of 1 GeV [17,18].\hfill\break\indent
To calculate the 1 loop diagrams shown in fig.1 we need the
couplings of the gluon to the gluinos,
of the scalar partners of the left handed quarks
to the gluon, photon and $Z^0$\ boson and of the gluino to the left
handed quark and its scalar partner.
The first one\footnote*{In order to shorten the notation we will use
$cos\Theta = c_\Theta,~sin\Theta = s_\Theta,~{\rm and}~s_W=sin\Theta_W$
where $\Theta_W$ is the weak mixing angle.}
 is given by Eq.C92 in [19]:
$${\cal L}_{g\tilde g\tilde g}={i\over 2}g_sf_{abc}
\overline{\tilde g_a}\gamma_\mu\tilde g_b G^\mu_c\eqno(3),$$
which is multiplied by 2 to obtain the
Feynman rules. The interactions of the gluon, photon
and the $Z^0$\ boson with squark are
given by Eq.6--8 in [9]:
$$\eqalignno{{\cal L}_{\tilde q\tilde q V}=&
-ieA^\mu\sum_{i=L,R}e_{q_i}\tilde q_i^\ast
{\buildrel \leftrightarrow \over\partial}_\mu\tilde q_i-
{{ie}\over{s_W~c_W}}Z^\mu
\sum_{i=L,R}\Bigr(T_{3q_i}-e_{q_i}s_W\Bigr)\tilde q_i^\ast
{\buildrel\leftrightarrow\over\partial}_\mu\tilde q_i&(4)\cr
&-ig_s T^a G^{a\mu}
\sum_{i=L,R}\tilde q_i^\ast
{\buildrel\leftrightarrow\over\partial}_\mu\tilde q_i
}$$
After the introduction of nontrivial squark mixing this becomes
$$\eqalignno{{\cal L}_{\tilde q\tilde q V}=&-ieA^\mu\sum_{i=1,2}
e_{q_i}\tilde q_i^\ast
{\buildrel\leftrightarrow\over\partial}_\mu\tilde q_i
-ig_s T^a G^{a\mu}\sum_{i=1,2}\tilde q_i^\ast
{\buildrel\leftrightarrow\over
\partial}_\mu\tilde q_i&(5)\cr
&-{{ie}\over{s_Wc_W}}Z^\mu\lbrack (T_{3L}c^2_\Theta-e_qs^2_W)
\tilde q_1^\ast
{\buildrel\leftrightarrow\over\partial}_\mu\tilde q_1
+(T_{3L}s^2_\Theta-e_qs^2_W)\tilde q_2^\ast
{\buildrel\leftrightarrow\over
\partial}_\mu\tilde q_2\cr &-T_{3L}c_\Theta s_\Theta(
\tilde q_1^\ast{\buildrel\leftrightarrow\over
\partial}_\mu\tilde q_2+
\tilde q_2^\ast{\buildrel\leftrightarrow\over
\partial}_\mu\tilde q_1)
\rbrack\cr}$$
Finally the coupling that leads to flavour changing is
given by Eq.1 in [16]:
$${\cal L_{FC}}=-\sqrt{2}g_sT^aK\overline{\tilde g}P_Lq(c_\Theta\tilde
q_1-s_\Theta\tilde q_2)+h.c.\eqno(6)$$
Here K is the supersymmetric version of the Kobayashi--Maskawa
matrix whose form will appear later. Flavour
changing couplings occur only in the left handed scalar
quark sector; the right handed sector does not contribute
to our process.\hfill\break\indent
\noindent
After summation over all diagrams, we obtain the following
effective $tcV$ vertex:
$$\eqalignno{M^\alpha_{\mu V}=&-i{{\alpha_s}\over{2\pi}}
K_{\alpha t}K_{\alpha c}\overline u_{p_2}\lbrack
\gamma_\mu P_L V_V^\alpha+{{P_\mu}\over{m_{\rm top}}}P_R
T_V^\alpha\rbrack u_{p_1}&(7)\cr
V_\gamma^\alpha=&ee_qC_2(F)\lbrack
c_{\Theta_\alpha}^2 (C^{11\alpha}_\epsilon-C^{1\alpha}_{SE})
+s_{\Theta_\alpha}^2(C^{22\alpha}_\epsilon-C^{2\alpha}_{SE})
\rbrack\cr
T_\gamma^\alpha=&ee_qC_2(F)\lbrack c_{\Theta_\alpha}^2
C^{11\alpha}_{\rm top}+s_{\Theta_\alpha}^2 C^{22\alpha}_{\rm top}
\rbrack\cr
V_g^\alpha=&g_sT^a\bigl\lbrace\lbrack -{1\over 2}C_2(G)+C_2(F)
\rbrack\lbrack c_{\Theta_\alpha}^2 C^{11\alpha}_\epsilon+
s_{\Theta_\alpha}^2 C^{22\alpha}_\epsilon\rbrack-C_2(F)\lbrack
c_{\Theta_\alpha}^2 C^{1\alpha}_{SE}+s_{\Theta_\alpha}^2
C^{2\alpha}_{SE}\rbrack\cr
&+{1\over 2}C_2(G)\lbrace c_{\Theta_\alpha}^2\lbrack
C^{\tilde g 1\alpha}_\epsilon+C^{1\alpha}_{\tilde g}
+C_{q^2}^{1\alpha}+C_t^{1\alpha}\rbrack
+s_{\Theta_\alpha}^2\lbrack
C^{\tilde g 2\alpha}_\epsilon+C^{2\alpha}_{\tilde g}
+C_{q^2}^{2\alpha}+C_t^{2\alpha}
\rbrack\rbrace\bigr\rbrace\cr
T_g^\alpha=&g_sT^a\bigl\lbrace\lbrack -{1\over 2}C_2(G)+C_2(F)
\rbrack\lbrack c_{\Theta_\alpha}^2 C^{11\alpha}_{\rm top}
+s_{\Theta_\alpha}^2 C^{22\alpha}_{\rm top}\rbrack
-{1\over 2}C_2(G)\lbrack c_{\Theta_\alpha}^2
C^{1\alpha}_t+s_{\Theta_\alpha}^2 C^{2\alpha}_t\rbrack
\bigr\rbrace\cr
V_Z^\alpha=&{{e}\over{s_Wc_W}}C_2(F)\lbrace
(T_{3L}c^2_{\Theta_\alpha}-e_qs_W^2)
c_{\Theta_\alpha}^2 C^{11\alpha}_\epsilon
+(T_{3L}s^2_{\Theta_\alpha}-e_qs_W^2)
s_{\Theta_\alpha}^2 C^{22\alpha}_\epsilon\cr
&+T_{3L}c^2_{\Theta_\alpha} s^2_{\Theta_\alpha}
(C^{12\alpha}_\epsilon+C^{21\alpha}_
\epsilon)-(T_{3L}-e_qs_W^2)\lbrack c^2_{\Theta_\alpha}
C^{1\alpha}_{SE}+s^2_{\Theta_\alpha} C^{2\alpha}_{SE}
\rbrack\rbrace\cr
T_Z^\alpha=&{{e}\over{s_Wc_W}}C_2(F)\lbrace
(T_{3L}c^2_{\Theta_\alpha}-e_qs_W^2)
c_{\Theta_\alpha}_2 C^{11\alpha}_{\rm top}
+(T_{3L}s^2_{\Theta_\alpha}-e_qs_W^2)
s_{\Theta_\alpha}^2 C^{22\alpha}_{\rm top}\cr
&+T_{3L}c^2_{\Theta_\alpha} s^2_{\Theta_\alpha}
(C^{12\alpha}_{\rm top}+C^{21\alpha}_{\rm top})\rbrace\cr}$$
$$\eqalignno{
C^{kl\alpha}_\epsilon
=&\int\limits_0^1d\alpha_1
\int\limits_0^{1-\alpha_1}d\alpha_2
\lbrack {1\over\epsilon}-\gamma+\ln(4\pi\mu^2)-\ln(f_{kl}^
\alpha)\rbrack\cr
 C^{kl\alpha}_{\rm top}=&
\int\limits_0^1d\alpha_1
\int\limits_0^{1-\alpha_1}d\alpha_2
{{m^2_{\rm top}\alpha_1(1-\alpha_1-\alpha_2)}\over{ f_{kl}^
\alpha}}\cr
C^{k\alpha}_{SE}=&\int\limits_0^1 d\alpha_1\alpha_1
\lbrack {1\over\epsilon}-\gamma+\ln(4\pi\mu^2)-\ln(g_k^\alpha)
\rbrack\cr
C^{\tilde g k\alpha}_\epsilon=&\int\limits_0^1d\alpha_1
\int\limits_0^{1-\alpha_1}d\alpha_2
\lbrack {1\over\epsilon}-\gamma+1+\ln(4\pi\mu^2)-\ln(h_k^
\alpha)\rbrack\cr
C^{k\alpha}_{\tilde g}=&\int\limits_0^1d\alpha_1
\int\limits_0^{1-\alpha_1}d\alpha_2 {{m_{\tilde g}^2}
\over{h^\alpha_k}}\cr
C_{q^2}^{k\alpha}=&\int\limits_0^1d\alpha_1
\int\limits_0^{1-\alpha_1}d\alpha_2{{q^2\alpha_1\alpha_2}\over
{h^\alpha_k}}\cr
C^{k\alpha}_t=&\int\limits_0^1d\alpha_1
\int\limits_0^{1-\alpha_1}d\alpha_2
{{m^2_{\rm top}\alpha_1(1-\alpha_1-\alpha_2)}\over{h_k^\alpha}}\cr
f_{kl}^{\alpha}=&m^2_{\tilde g}-(m^2_{\tilde g}-
m_{\tilde q^\alpha_k}^2)\alpha_1-(m^2_{\tilde g}-
m_{\tilde q^\alpha_l}^2)\alpha_2-
m^2_{\rm top}\alpha_1(1-\alpha_1-\alpha_2)
- q^2\alpha_1\alpha_2\cr
g^\alpha_k=&m^2_{\tilde g}-(m^2_{\tilde g}-
m_{\tilde q^\alpha_k}^2)\alpha_1-
m^2_{\rm top}\alpha_1(1-\alpha_1)\cr
h^\alpha_k=&m_{\tilde q^\alpha_k}^2-(m_{\tilde q^\alpha_k}^2-
m^2_{\tilde g})(\alpha_1+\alpha_2)-
m^2_{\rm top}\alpha_1(1-\alpha_1-\alpha_2)
- q^2\alpha_1\alpha_2\cr}
$$
where $\epsilon = 2-d/2$,\ $C_2(F)=4/3$\ and $C_2(G)=3$\ for SU(3).
If
$\alpha\not={\rm top}$\ we have $c_{\Theta_\alpha}=1$.
Using the spin-1 condition ($q_\mu=(p_1-p_2)_\mu=0$)\ we
can write $P_\mu=(p_1+p_2)_\mu=2p_{1\mu}$.
$K_{\alpha q}$\ is the SUSY--Kobayashi--Maskawa matrix whose form
is as follows:
$$K_{ij}=\left(\matrix{1&\varepsilon&\varepsilon^2\cr
 -\varepsilon&1&\varepsilon\cr-\varepsilon^2&-\varepsilon&1\cr}\right)
\eqno (8)$$
Here $\varepsilon$\ is a small number
(not to be confused with the $\epsilon$\ above) to be taken as
$\varepsilon^2=1/4$\ [16,8].
It is straightforward at this point to verify that all divergent terms cancel
exactly, without the GIM mechanism.
\hfill\break\indent
A crucial test is also provided by the nature of the current.
Using the following identity:
$$\overline u_{p_2}{{P^\mu}\over{m_{\rm top}}}P_R u_{p_1}
\equiv \overline u_{p_2}\lbrack\gamma_\mu P_L+
i\sigma_{\mu\nu}{{q^\mu}\over{m_{\rm top}}}P_R\rbrack u_{p_1}
\eqno(9)$$
we can show that the quantity in front of the $\gamma^\mu$ term vanishes in the
limit $q^2\to 0$, as required by gauge invariance.
\hfil\break\indent
%$$C^{ii\alpha}_\epsilon+C^{ii\alpha}_{\rm top}-C^{i\alpha}_
%{SE}\vert_{q^2=0}\equiv 0\equiv
% C^{ii\alpha}_\epsilon+C^{ii\alpha}_{\rm top}
%-C^{\tilde g i\alpha}_\epsilon-C^{i\alpha}_{\tilde g}\vert_
%{q^2=0}\eqno(10)$$
When summing over all scalar quarks within the loops
the scalar up quark cancels out because of the unitarity
of $K_{ij}$\ and with $K_{23}=-K_{32}$\ the mass
splitting of the scalar top quark and the scalar charm
quark comes into account, which was taken
to be $m_{\tilde c}=0.9~m_{\tilde t}$\ in [8] and therefore
too small for a top quark mass of 174 GeV.
If all scalar quark masses would be the same the
decay rate of $t\rightarrow cV$\ would be identical
to 0. As a final result we obtain:
$$\eqalignno{\Gamma_S(t\rightarrow c V)=&
%({1\over 2}-{2\over 3}\sin^2\Theta_W)^2{{\alpha\alpha_s^2}
{{\alpha_s^2}
\over{128\pi^3}}~m_{\rm top}~
\biggl( 1-{{m_{V}^2}\over{m^2_{\rm top}}} \biggr)^2\varepsilon^2\cr
\Bigl\lbrack V_V^2\Bigl( 2&+{{m^2_{\rm top}}\over{m^2_{V}}} \Bigr)-
2V_VT_V\Bigl( 1-{{m^2_{\rm top}}\over{m^2_{V}}} \Bigr)-
T_V^2\Bigl( 2-{{m^2_{V}}\over{m^2_{\rm top}}}-
{{m^2_{\rm top}}\over{m^2_{V}}} \Bigr) \Bigr\rbrack&(11)\cr}$$
\noindent
where  $\displaystyle{V_V=V_V^{\tilde t}-V_V^{\tilde c}}$\
and $\displaystyle{T_V=T_V^{\tilde t}-T_V^{\tilde c}}$.
\noindent
For $V=\gamma,g$ we have $V_V=-T_V$\ and
all terms containing $m_V^2$\ are absent.\hfill\break
\noindent
We define [6]:
$\displaystyle{B(t\rightarrow cV)=\Gamma_S(t\rightarrow cV)/
\Gamma_W(t\rightarrow bW^+)}$\ where
$$\Gamma_W(t\rightarrow bW^+)={{\alpha}\over{16\sin^2\Theta_W}}
m_{\rm top}\Bigl(1-{{m^2_{W^+}}\over{m^2_{\rm top}}}\Bigr)^2\Bigl(2+
{{m^2_{\rm top}}\over{m^2_{W^+}}}\Bigr)\eqno(12)$$
Our input parameters are
$m_{\rm top}=174$\ GeV and the strong coupling
constant $\alpha_s=1.4675/\ln({{m^2_{\rm top}}\over
{\Lambda^2_{\rm QCD}}})=0.107$\ with $\Lambda_{\rm QCD}=
0.18$\ GeV [6].\hfill\break\indent
In fig. 2 we present the branching ratio
$B(t\rightarrow c Z)$
\ as a function of the scalar
mass $m_S$\ for a gluino mass of 100 GeV.
We see that without mixing, the branching
ratio decreases rapidly with increasing scalar mass.
The mixing has a drastic effect. It enhances the branching ratio
 by up to 5 orders of
magnitude for large $m_s$.
Higher values of $\tan\beta$\ diminish the branching
ratio. The gluino mass hardly affects the decay rate.
Even for a small gluino mass of the order of 1 GeV the
branching ratio remains of the same order.
\hfill\break\indent
In fig. 3 we consider the same cases as
in fig. 2 but for $B(t\rightarrow c g)$.
The effect of the mixing is not as drastic as in the
previous case. It decreases the branching ratio
generally by 1--2 orders of magnitude. This reduction
is larger for larger scalar masses.
Increasing $\tan\beta$\ diminishes the branching ratio
in general, an exception is the case $\mu=100$\ GeV
and $m_{\tilde g}=500$\ GeV. Increasing the gluino mass
diminishes the branching ratio by several orders of magnitude
for lower values of the scalar mass whereas
lower values of the gluino mass enhances the ratio.
The shape of the figures remains the same.
\hfill\break\indent
In figs. 4 and 5  we consider the branching ratio $B(t\rightarrow c \gamma)$
We notice first that the effect of the mixing
is rather small for small values of $m_s$. We also note that the sensitivity
of the branching ratio to $\tan\beta$ is greatly increased. Thirdly, one
sees that the mixing generally reduces the branching ratio. This
is true generally but might not hold for some regions of parameter
space, as can be seen on fig. 4, when some combinations of parameters
can greatly increase the branching ratio.
Most interesting, the mixing gives rise to a GIM-like suppression where the
contribution of the top quark exactly cancels the contribution from the
c-quark. This dramatic cancellation is also seen on fig. 4.
Such a cancellation is not {\it isolated} as seen on
fig. 5\footnote*{This figure is intended to give a very good idea of the global
behaviour but not to be read numerically.}.
We have tried many different combinations of $\mu$ and
$m_{\tilde g}$ and we found a {\it rift} similar to the one visible on
fig. 5 with all the combinations. Such a cancellation does not occur for the
gluon and Z decay modes. In the first case, the $g-\tilde g-\tilde g$ vertex
spoils it while in the second case it seems to be the $q^2\ne 0$ that does
it.
In this paper we presented the supersymmetric QCD 1 loop correction
to the
flavour changing decay rate $t\rightarrow cV$.
We have shown that the $t\rightarrow c Z$\ decay rate
is enhanced by several orders of magnitude compared
to the standard model. If we include the mixing of
the scalar partners of the top quark we do get a
further enhancement and the decay rate
remains relatively large for a very wide range of gluino and
scalar masses.
For the $t\rightarrow c g$\ decay rate we have shown
that the mixing reduces generally the branching ratio.
Larger values for $\tan\beta$\ also diminish the branching
ratio.
In the $t\rightarrow c
\gamma$ decay mode, the most dramatic effect of this mixing is to give rise
to a GIM-like cancellation for some combinations of parameters. It also reduces
the branching ratio and greatly increases the sensitivity to $\tan\beta$.
\hfill\break\vskip.1cm\noindent
One of us (H.K.) would like to thank the physics department
of Carleton university and Universit\'e de Montr\'eal
for the use of their computer
facilities as well as M. Boyce for computerial advices.
The figures were done with the very user
friendly program PlotData from TRIUMF.
This work was partially funded by N.S.E.R.C. of
Canada and Les Fonds F.C.A.R. du Qu\'ebec.
\hfill\break\vskip.1cm\noindent
\item{[\ 1]}CDF Collaboration, Fermilab preprint, April 1994.
\item{[\ 2]}J.L. Diaz-Cruz et al, Phys.Rev.{\bf D41}
(1990)891.
\item{[\ 3]} B. Dutta Roy et al, Phys.Rev.Lett {\bf 65}
(1990)827.
\item{[\ 4]}H. Fritzsch, Phys.Lett.{\bf B224}(1989)423.
\item{[\ 5]}W. Buchm\"uller and M. Gronau, Phys.Lett.{
\bf B220}(1989)641.
\item{[\ 6]}G. Eilam, J.L. Hewett and A. Soni, Phys. Rev.
{\bf D44}(1991)1473
\item{[\ 7]}B. Grzadkowski, J.F. Gunion and P. Krawczyk,
Phys.Lett.{\bf B268}(1991)106.
\item{[\ 8]}C.S. Li, R.J. Oakes and J.M. Yang, Phys.Rev.
{\bf D49}(1994)293.
\item{[\ 9]} A. Djouadi, M.Drees and H. K\"onig, Phys.Rev.
{\bf D48}(1993)3081.
\item{[10]}J. Ellis and D.V. Nanopoulos, Phys.Lett{
\bf 110B}(1982)44.
\item{[11]}R. Barbieri and R. Gatto, Phys.Lett{\bf 110B}
(1982)211.
\item{[12]}T. Inami and C.S. Lim, Nucl.Phys.{\bf B207}
(1982)533.
\item{[13]}B.A. Campbell, Phys. Rev.{\bf D28}(1983)209.
\item{[14]}M.J. Duncan, Nucl.Phys.{\bf B221}(1983)221.
\item{[15]}J.F. Donoghue, H.P. Nilles and D. Wyler,
Phys.Lett.{\bf 128 B}(1983)55.
\item{[16]}M.J. Duncan, Phys.Rev.{\bf D31}(1985)1139.
\item{[17]} see e.g. HELIOS collaboration, T. Akesson et al,
Z.Phys.{\bf C52}(1991)219 and references therein.
\item{[18]}J. Ellis, D.V. Nanopoulos and D.A. Ross, Phys.Lett.
{\bf B305}(1993)375.
\item{[19]}H.E. Haber and G.L. Kane, Phys.Rep.{\bf 117}(1985)75.
%\vfill\break
\hfill\break\vskip.12cm\noindent
%\vfill\break\noindent
{\bf FIGURE CAPTIONS}\vskip.12cm
\item{Fig.1} The diagrams with scalar quarks and
gluinos within the loop, which contribute to
the top quark decay into a charm quark and
a Z boson, photon or gluon.
\item{Fig.2}The ratio $\Gamma_S/\Gamma_W$\ of the
the top quark decay
into a charm quark and $Z^0$\ boson as a function of the
scalar mass $m_S$. The gluino mass was taken to be
100 GeV.
The solid line is
the unphysical case with no mixing ($\mu=0=A_{\rm top}$) and
$\tan\beta=1$, the dotted line the same case with
$\tan\beta=10$. The other cases are with mixing
($A_{\rm top}=m_S$).
The dashed lines are with $\mu=100$\ GeV and the
dashed--dotted ones with $\mu=500$\ GeV. The shorter ones
are with $\tan\beta=1$\ and the longer ones with
$\tan\beta=10$.
%mass of 3 GeV (small mass window).
\item{Fig.3} The same as Fig.2  but for the
decay of the top quark into a charm quark and a gluon.
\item{Fig.4} The same as in Fig.2 but for the decay of the
top quark into a charm quark and a photon. The solid line with a
sharp dip corresponds to $\mu = 100~GeV$ and $\tan\beta = 2$.
\item{Fig.5} $Log_{10}(t\to c\gamma)$ as a function of $m_s$ and $tan\beta$
for $m_{\tilde g} = 100~GeV = \mu$. The vertical scale is about the same as on
fig. 4.
\vfill\break
\end